\documentclass[aps,pra,twocolumn,showpacs,prabib]{revtex4}
\usepackage{graphicx}
\usepackage{amsmath}
\usepackage{amssymb,wasysym}
\usepackage{amsfonts}
\usepackage{bm,verbatim}

\newcommand{\bea}{\begin{eqnarray}}
\newcommand{\eea}{\end{eqnarray}}
\newcommand{\be}{\begin{equation}}
\newcommand{\ee}{\end{equation}}

\newcommand{\rr}{\mathbf{r}}
\newcommand{\kk}{\mathbf{k}}

\newcommand{\qq}{\mathbf{q}}

\newcommand{\rhob}{\mbox{\boldmath$\rho$}}
\newcommand{\rhobs}{\mbox{\scriptsize\boldmath$\rho$}}

\begin{document}

\title{Single-Particle Momentum Distribution of an Efimov trimer}

\author{Yvan Castin}
\affiliation{Laboratoire Kastler Brossel, \'Ecole Normale
Sup\'erieure, UPMC and CNRS, 24 rue Lhomond, 75231 Paris Cedex 05, France}

\author{F\'elix Werner}
\affiliation{Department of Physics, University of Massachusetts,
Amherst, MA 01003, USA}

\begin{abstract}
Experimental progress in the study of strongly interacting
ultracold atoms has recently allowed the observation of Efimov trimers.
We study theoretically 
a non-conventional observable for these trimer states, that may be accessed
experimentally,
the momentum distribution $n(\kk)$ of the constitutive bosonic particles.
The large momentum part of the distribution is particularly intriguing:
In addition to the expected $1/k^4$ tail associated to contact interactions,
it exhibits a subleading tail $1/k^5$ which is a hall-mark of Efimov
physics and leads to a breakdown of a previously proposed expression 
of the energy as a functional of the momentum distribution.
\end{abstract}

\pacs{31.15.ac,03.75.Hh}

\maketitle
\section{Introduction}

Experiments with ultracold atoms have now entered the regime of strong
interactions, thanks to the possibility to manipulate the $s$-wave
scattering length $a$ between cold atoms with a magnetically induced Feshbach
resonance \cite{Feshbach,revueStringariFermions}. 
This has led to a revolution in the study of the few-body problem,
as one can now have in a controllable way
a scattering length much larger (in absolute value)
than the range $b$ (and the effective range) of the interaction potential.
In particular, this has allowed to confirm experimentally 
\cite{Efimov_manips,Jochim} 
the existence of the long-searched Efimov effect \cite{BH,Efimov1,Efimov2}:
As shown by Efimov in the early 1970's, three particles interacting
{\sl via} a short range potential with an infinite scattering length may exhibit
an infinite number of trimer states with 
a geometric spectrum. The existence of an infinite number of bound states 
is usual, even at the two-body level, for long range interactions,
but it is quite intriguing for short range interaction potentials. 
This Efimov effect takes place for three (same spin state) bosons \cite{Efimov1}, 
but it is more general,
it also occurs for example for two (same spin state) fermions and a
third distinguishable particle at least $13.607$ times lighter
\cite{Efimov2,PetrovEfim}.

On the experimental side, an increasing number of observable quantities
are now at hand. For Efimov physics, the usual evidence of the emergence
of an Efimov trimer state is a peak in the three-body loss rate as a function 
of the scattering length \cite{Efimov_manips}. 
Now radio-frequency spectroscopic
techniques can give a direct access to the trimer spectrum
\cite{Jochim}.
For strongly interacting Fermi gases (without Efimov effect) 
a very precise measurement of the atomic momentum distribution 
$n(\kk)$ was performed recently, so precise that it allowed to
see the large momentum tail $n(\kk)\sim C/k^4$, $k$ large but still smaller
than $1/b$, and to quantitatively extract the coefficient $C$ whose values
were satisfactorily compared to theory \cite{Jin_nk}.
The same conclusion holds for the few-body numerical experiment of \cite{Blume}.
Similarly, the first order coherence function $g^{(1)}$ of the atomic field
over a distance $r$, a quantity measured for bosonic cold atoms \cite{Esslinger}
but not yet for fermionic cold atoms,
is related to the Fourier transform of $n(\kk)$ and is sensitive 
to the $1/k^4$ tail by a contribution that is non-differentiable
with respect to the {\sl vector} $\rr$ in $\rr=\mathbf{0}$ \cite{lien_g1}, and that appeared
in the many-body numerical experiment of \cite{Astra}.

The occurrence of the $1/k^4$ tail in $n(\kk)$ is a direct consequence of two-body
physics, that is of the binary
zero-range interaction between two particles, and it holds in all spatial
dimensions: According to Schr\"odinger's equation for the zero-energy
scattering
state $\phi(\rr)$ of two particles of relative coordinates
$\rr$, $\Delta_{\rr} \phi(\rr)\propto \delta(\rr)$ for a contact
(regularized Dirac delta) interaction \cite{contact_interaction}, 
so that in Fourier space $\tilde{\phi}(\kk)
\propto 1/k^2$ and $n(\kk)\propto |\tilde{\phi}(\kk)|^2$ scales
as $1/k^4$. On the contrary, the coefficient $C$, called contact,
depends on the many-body properties,
and can be related to the derivative of the gas mean energy (or mean free
energy at non-zero temperature) with respect to the scattering length,
as was shown first for bosons in one dimension \cite{Olshanii},
then for spin 1/2 fermions in three dimensions \cite{Tan1,Braaten_contact,Tarruell}, 
and for bosonic or fermionic, three dimensional or bidimensional systems,
in \cite{tangen}. 

In this paper, we anticipate that experimentally, it may be possible
to measure with high precision the atomic momentum distribution 
in systems subjected to the Efimov effect, for example
in a Bose gas with a large scattering length \cite{HammerBosons,Cornell_Bose,Hulet,Chevy}. 
To be specific, we consider in the center of mass frame
the Efimov trimer states for three bosons interacting with infinite
scattering length. After recalling the expression of the three-body
wavefunction in section \ref{sec:wf}, we obtain the expression
of the momentum distribution in terms of integrals over a single momentum vector
in section \ref{sec:nk}, see Eqs.(\ref{eq:decomp},\ref{eq:nI},\ref{eq:nII},\ref{eq:nIII},\ref{eq:nIV}).
As illustrated in section \ref{sec:appli}, this allows
to perform a very precise numerical evaluation
of the momentum distribution for all values of the single-particle wavevector
$\kk$, and to analytically obtain the large momentum behavior of
$n(\kk)$: In addition to the expected $C/k^4$ term at large $k$, we find
an unexpected $1/k^5$ subleading term, that is a direct and generic
signature of Efimov physics \cite{cluster_Tan}, see (\ref{eq:nk_efi}). Another, more formal, 
consequence of this $1/k^5$
subleading term is that the general expression giving the 
energy as a functional of the momentum distribution $n(\kk)$,
derived for the non-Efimovian case in \cite{Tan2} and extended (with the
same form) to the Efimovian case in \cite{Leyronas}, 
turns out to be invalid in the Efimovian case \cite{note2D}.
We conclude in section \ref{sec:conclusion}.

\section{Normalized wavefunction of an Efimov trimer}
\label{sec:wf}

\subsection{Three-body state in position space}

In this subsection we recall the  wavefunction of an Efimov trimer and give the expression of its normalization constant.
We consider an Efimov trimer state for three same-spin-state bosons of mass $m$ interacting {\sl via} a zero
range potential with infinite scattering length.
In order to avoid formal normalisability problems, it is convenient to imagine that the Efimov trimer
is trapped in an arbitrarily weak harmonic potential, that is with a ground state
harmonic oscillator length  $a_{\rm ho}$ arbitrarily larger than the trimer size \cite{general}.
Since the center of mass of the system is separable in a harmonic potential,
this fixes the normalisability problem without affecting the internal wavefunction
of the trimer in the limit $a_{\rm ho}\to +\infty$.
In this limit, the energy of the trimer is equal to the free space energy
\be
E_{\rm trim} = - \frac{\hbar^2 \kappa_0^2}{m}, \ \ \ \kappa_0>0.
\ee
According to Efimov's asymptotic, zero-range theory \cite{Efimov1}, 
\be
\label{eq:kappa0}
\kappa_0 = \frac{\sqrt{2}}{R_t} e^{-\pi q/|s_0|} e^{\mbox{\scriptsize Arg}\,\Gamma(1+s_0)/|s_0|}
\ee
where $R_t>0$ is a length known as the three-body parameter \cite{def_Rt},  the quantum number
$q$ is any integer in $\mathbb{Z}$,
and the purely imaginary number $s_0 = i |s_0|$
is such that
\be
\label{eq:def_ms0}
|s_0| \cosh(\frac{|s_0|\pi}{2}) = \frac{8}{\sqrt{3}} \sinh(\frac{|s_0|\pi}{6}),
\ee
so that $|s_0|=1.00623782510\ldots$
The corresponding three-body wavefunction $\Psi$ may be written 
for $\kappa_0 a_{\rm ho}\to +\infty$ as
\begin{multline}
\label{eq:etat}
\Psi(\mathbf{r}_1,\mathbf{r}_2,\mathbf{r}_3) 
=\psi_{\rm CM}(\mathbf{C}) \left[
\psi(r_{12},\frac{|2\mathbf{r}_3-(\mathbf{r}_1+\mathbf{r}_2)|}{\sqrt{3}})\right. \\
\left. +\psi(r_{23},\frac{|2\mathbf{r}_1-(\mathbf{r}_2+\mathbf{r}_3)|}{\sqrt{3}})
+ \psi(r_{31},\frac{|2\mathbf{r}_2-(\mathbf{r}_3+\mathbf{r}_1)|}{\sqrt{3}})
\right],
\end{multline}
where $\mathbf{C}=(\mathbf{r}_1+\mathbf{r}_2+\mathbf{r}_3)/3$ is the center of mass
position of the three particles and the parameterization of $\psi$
is related to the Jacobi coordinates 
$\mathbf{r}=\mathbf{r}_2-\mathbf{r}_1$
and $\rhob=[2\mathbf{r}_3-(\mathbf{r}_1+\mathbf{r}_2)]/\sqrt{3}$.
In our expression of $\Psi$, $\psi_{\rm CM}$ is the Gaussian wavefunction of the center of mass ground state in the harmonic
trap, normalized to unity, and $\psi$ is a Faddeev component of the free space trimer wavefunction.
The explicit expression of $\psi$ is known \cite{Efimov1}:
\be
\label{eq:psi_F}
\psi(r,\rho)  = \frac{\mathcal{N}_\psi}{\sqrt{4\pi}}  
\frac{K_{s_0}(\kappa_0\sqrt{r^2+\rho^2})}{(r^2+\rho^2)/2}
\frac{\sin[s_0(\frac{\pi}{2}-\alpha)]}{\sin(2\alpha)}
\ee
where $K_{s_0}$ is a Bessel function and $\alpha=\mbox{atan}(r/\rho)$.
The normalization factor $\mathcal{N}_\psi$ ensuring that $||\Psi||^2=1$ was not calculated 
in \cite{Efimov1}. To obtain its explicit expression,
one first
performs the change of variables
$(\mathbf{r}_1,\mathbf{r}_2,\mathbf{r}_3)\to (\mathbf{C}, \rhob, \mathbf{r})$, whose Jacobian
is  $D(\rr_1,\rr_2,\rr_3)/D(\mathbf{C},\rhob,\rr)=(-\sqrt{3}/2)^3$. To integrate over $\mathbf{r}$ and $\rhob$ one
then introduces hyperspherical coordinates in which the wavefunction separates; one then faces known
integrals on the hyperradius \cite{Gradstein} and on the hyperangles \cite{Efimov93}.
This leads to \cite{WernerThese}:
\begin{multline}
\label{eq:norma}
|\mathcal{N}_\psi|^{-2} =
\left(\frac{\sqrt{3}}{2}\right)^3
\frac{3\pi^2}{2\kappa_0^2 \cosh(\frac{|s_0|\pi}{2})} \\
\times \left[\cosh(\frac{|s_0|\pi}{2}) +\frac{|s_0|\pi}{2} \sinh(\frac{|s_0|\pi}{2}) 
-\frac{4\pi}{3\sqrt{3}}\cosh(\frac{|s_0|\pi}{6})\right].
\end{multline}

\subsection{Three-body state in momentum space}

To obtain the momentum distribution for the Efimov trimer, we need to evaluate the Fourier transform
of the trimer wavefunction $\Psi$ given by (\ref{eq:etat}). 
Rather than directly using (\ref{eq:psi_F}), we take advantage of the fact that, for contact interactions,
the Faddeev component $\psi$ obeys the non-interacting Schr\"odinger's equation 
with a source term. With the change to Jacobi coordinates,
the Laplace operator in the coordinate space of dimension nine
reads
$\sum_{i=1}^{3} \Delta_{\mathbf{r}_i} = \frac{1}{3}\Delta_{\mathbf{C}}
+2\left[\Delta_{\mathbf{r}}+\Delta_{\rhob}\right]$ so that 
\be
\label{eq:Schr}
-\left[\kappa_0^2-\Delta_{\mathbf{r}}-\Delta_{\rhobs}\right]
\psi(r,\rho) = \delta(\mathbf{r}) B(\rho).
\ee
The source term in the right hand side originates from the fact that 
\be
\psi(r,\rho) \underset{r\to 0}{\sim} -\frac{B(\rho)}{4\pi r}
\ee
for a fixed $\rho$, this $1/r$ divergence coming from the replacement
of the interaction potential by the Bethe-Peierls contact condition.
Taking the Fourier transform of (\ref{eq:Schr}) over $\mathbf{r}$ and
$\rhob$ leads to
\be
\label{eq:psit}
\tilde{\psi}(k,K) = -\frac{\tilde{B}(K)}{k^2+K^2+\kappa_0^2},
\ee
where the Fourier transform is defined as 
$\tilde{B}(K) \equiv \int d^3\rho\,  e^{-i\mathbf{K}\cdot\rhobs} B(\rho)$.
$B(\rho)$ is readily obtained from (\ref{eq:psi_F}) by taking the limit $r\to 0$:
\be
B(\rho) = -\mathcal{N}_\psi (4\pi)^{1/2} i \sinh(\frac{|s_0|\pi}{2}) \frac{K_{s_0}(\kappa_0\rho)}{\rho}.
\ee
The Fourier transform of this expression is known, see relation 6.671(5)
in \cite{Gradstein}, so that
\begin{multline}
\label{eq:Bt}
\!\!\!\!\!\! \tilde{B}(K) = \frac{-2\pi^{5/2}\mathcal{N}_\psi}{K (K^2+\kappa_0^2)^{1/2}}
\left\{
\left[\frac{(K^2+\kappa_0^2)^{1/2}+K}{\kappa_0}\right]^{s_0}\!\!\!\!-\mbox{c.c.}\right\}
\end{multline}
where $\mbox{c.c.}$ stands for the complex conjugate.
Note that the expression between the curly brackets simply reduces to $2i\sin(|s_0|\alpha)$
if one sets $K=\kappa_0 \sinh \alpha$.
What we shall need is the large $K$ behavior of $\tilde{B}(K)$. Expanding (\ref{eq:Bt})
in powers of $\kappa_0/K$ gives
\be
\label{eq:agk}
\tilde{B}(K) = \mathcal{N}_\psi \frac{2\pi^{5/2}}{K^2}
\left[(2K/\kappa_0)^{-s_0}-\mbox{c.c.}\right]
+O(1/K^4).
\ee

The last step to obtain the trimer state vector in momentum space
is to take the Fourier transform of (\ref{eq:etat}), using the appropriate Jacobi coordinates
for each Faddeev component, or simply by Fourier transforming the first Faddeev component
using the coordinates $(\mathbf{C},\mathbf{r},\rhob)$ given above
and by performing circular permutations on the particle labels. This gives
\begin{multline}
\label{eq:etatf}
\tilde{\Psi}(\kk_1,\kk_2,\kk_3)=\left(\frac{\sqrt{3}}{2}\right)^3
\tilde{\psi}_{\rm CM}(\kk_1+\kk_2+\kk_3)\\
\times\Big[
\tilde{\psi}\left(\frac{|\kk_2-\kk_1|}{2},
\frac{|2\kk_3-(\kk_1+\kk_2)|}{2\sqrt{3}}\right)
+(231)+ (312)\Big],
\end{multline}
where the notation $(ijk)$ means that the indices $1,2,3$ have been replaced by
$i,j,k$ respectively.

\section{Integral expression of the momentum distribution}
\label{sec:nk}

To obtain the momentum distribution for an Efimov trimer state, 
it remains to integrate over
$\kk_3$ and $\kk_2$ the modulus square of the Fourier transform
(\ref{eq:etatf}) of the trimer wavefunction.
In the limit $\kappa_0 a_{\rm ho}\to +\infty$ where one suppresses the
harmonic trapping, one can set
\be
|\tilde{\psi}_{\rm CM}(\kk_1+\kk_2+\kk_3)|^2 = (2\pi)^3 \delta(\kk_1+\kk_2+\kk_3)
\ee
so that the trimer is at rest in all what follows.
Integration over $\kk_3$ is then straightforward:
\begin{multline}
\label{eq:nk_long}
n(\kk_1) = 
3\left(\frac{\sqrt{3}}{2}\right)^6 
\int\frac{d^3 k_2}{(2\pi)^3} 
\Big|
\tilde{\psi}(\frac{|\kk_2-\kk_1|}{2},\frac{\sqrt{3}|\kk_1+\kk_2|}{2}) \\
+
\tilde{\psi}(|\kk_2+\frac{1}{2}\kk_1|,\frac{\sqrt{3}}{2}k_1)
+
\tilde{\psi}(|\kk_1+\frac{1}{2}\kk_2|,\frac{\sqrt{3}}{2}k_2)
\Big|^2.
\end{multline}
The factor $3$ in the right hand side results from the fact that,
as e.g. in \cite{Tan1}, we normalize the momentum distribution $n(\kk)$
to the total number of particles (rather than to unity):
\be
\int \frac{d^3k}{(2\pi)^3} \, n(\kk) = 3.
\ee
Also note that the sum of the squares of the 
arguments of $\tilde{\psi}$ is constant and equal to $k_1^2+k_2^2+\kk_1\cdot\kk_2$
for each term in the right hand side of (\ref{eq:nk_long}).
When using (\ref{eq:psit}), one can thus put the denominator in (\ref{eq:psit}) as a common denominator,
to obtain
\begin{multline}
\label{eq:nkb}
\!\!\!\!\!\! n(\kk_1) =  \!\!
\int\!\! \frac{d^3 k_2}{(2\pi)^3}
\frac{\left[\tilde{B}(\frac{\sqrt{3}}{2}|\kk_1+\kk_2|)
+\tilde{B}(\frac{\sqrt{3}}{2}k_1) + \tilde{B}(\frac{\sqrt{3}}{2}k_2)
\right]^2}{(4^3/3^4)(k_1^2+k_2^2+\kk_1\cdot\kk_2+\kappa_0^2)^2}.
\end{multline}
For simplicity, we have assumed that the normalization factor
$\mathcal{N}_\psi$ is purely imaginary, so that $\tilde{B}(K)$ is a
real quantity.

In the above writing of $n(\kk_1)$, 
the only ``nasty" contribution is $\tilde{B}(\sqrt{3}|\kk_1+\kk_2|/2)$;
the other contributions are ``nice" since they only depend on the moduli $k_1$ and $k_2$.
Expanding the square in the numerator of (\ref{eq:nkb}), one gets six terms, three squared terms
and three crossed terms.
The change of variable $\kk_2=-(\kk'_2+\kk_1)$ allows, in one of the 
squared terms and in one
of the crossed terms, to transform a nasty term into a nice term.
What remains is a nasty crossed term that cannot be turned into a nice one; in that term,
as  a compromise, one performs the change of variable $\kk_2=-(\kk_2'+\kk_1/2)$.
We finally obtain the momentum distribution as the sum of four contributions,
\be
\label{eq:decomp}
n(\kk_1) = n_{\rm I}(\kk_1) + n_{\rm II}(\kk_1) + n_{\rm III}(\kk_1) + n_{\rm IV}(\kk_1),
\ee
with
\bea
\label{eq:nI}
n_{\rm I}(\kk_1) &\!\!\! =\!\! &  
\frac{3^4}{4^3}\!\!  \int\!\! \frac{d^3 k_2}{(2\pi)^3}
\frac{\tilde{B}^2(\frac{\sqrt{3}}{2}k_1)}{(k_1^2+k_2^2+\kk_1\cdot\kk_2+\kappa_0^2)^2} 
\\
\label{eq:nII}
n_{\rm II}(\kk_1) &\!\!\! =\!\! & 
\frac{3^4}{4^3}\!\!  \int\!\! \frac{d^3 k_2}{(2\pi)^3}
\frac{2\tilde{B}^2(\frac{\sqrt{3}}{2}k_2)}{(k_1^2+k_2^2+\kk_1\cdot\kk_2+\kappa_0^2)^2} \\
\label{eq:nIII}
n_{\rm III}(\kk_1) &\!\!\! =\!\! & 
\frac{3^4}{4^3}\!\!  \int\!\! \frac{d^3 k_2}{(2\pi)^3}
\frac{4 \tilde{B}(\frac{\sqrt{3}}{2}k_1) \tilde{B}(\frac{\sqrt{3}}{2}k_2)}
{(k_1^2+k_2^2+\kk_1\cdot\kk_2+\kappa_0^2)^2} \\
\label{eq:nIV}
n_{\rm IV}(\kk_1) &\!\!\! =\!\! & 
\frac{3^4}{4^3}\!\!  \int\!\! \frac{d^3 k_2}{(2\pi)^3}
\frac{2\tilde{B}(\frac{\sqrt{3}}{4}|2\kk_2+\kk_1|)\tilde{B}(\frac{\sqrt{3}}{4}|2\kk_2-\kk_1|)}
{(\kappa_0^2+k_2^2+\frac{3}{4}k_1^2)^2}. \nonumber \\
&&
\eea

An interesting question is to know if one can go beyond the integral
expressions Eqs.(\ref{eq:nI},\ref{eq:nII},\ref{eq:nIII},\ref{eq:nIV}), that is
if one can obtain an explicit expression for the momentum distribution, at most in terms of
special functions.
The contribution $n_{\rm I}(\kk_1)$ is straightforward to calculate:
\be
\label{eq:nI_expl}
n_{\rm I}(\kk_1) = \frac{\sqrt{3}}{4\pi}
\left(\frac{3}{4}\right)^{3}
\frac{\tilde{B}^2(\sqrt{3}k_1/2)}{(k_1^2+4\kappa_0^2/3)^{1/2}}.
\ee
The contribution $n_{\rm II}(\kk_1)$ is also explicitly calculable
by performing the change of variable
$k_2=(2/\sqrt{3})\kappa_0\sinh\alpha$ and using the identity that can be derived from contour 
integration:
\be
\int_{-\infty}^{+\infty} d\alpha\, \frac{e^{is\alpha}}{\cosh\alpha 
-\cosh\alpha_0}
= \frac{2\pi\sin[s(i\pi-\alpha_0)]}{\sinh\alpha_0 \sinh(s\pi)},
\ee
where $s$ is any real number and $\alpha_0$ is any complex number with $0<\mbox{Im}\, \alpha_0 < \pi$.
This also allows to obtain an explicit expression of $n_{\rm III}(\kk_1)$
if one further applies integration by part, integrating the
factor $\sin(|s_0|\alpha)$. 
We do not give however the resulting expressions since,
contrarily to these first three contributions to $n(\kk_1)$, the contribution
$n_{\rm IV}(\kk_1)$ in (\ref{eq:nIV}) blocked our attempt to
calculate $n(\kk_1)$ explicitly. For $\kk_1=\mathbf{0}$
however it becomes equal to the contribution $n_{\rm II}$.
$n(\kk_1=0)$ can thus be evaluated explicitly in terms of $\kappa_0$ and $s_0$,
see Appendix F in \cite{tangen}. In numerical form this gives
\be
\label{eq:n_ori}
n(\kk=\mathbf{0}) =  \frac{55.43379775608\ldots}{\kappa_0^3}.
\ee

\section{Applications}
\label{sec:appli}

\subsection{Numerical evaluation of $n(\kk)$ at all $k$} 

\begin{figure*}[htb]
\includegraphics[width=0.445\linewidth,clip=]{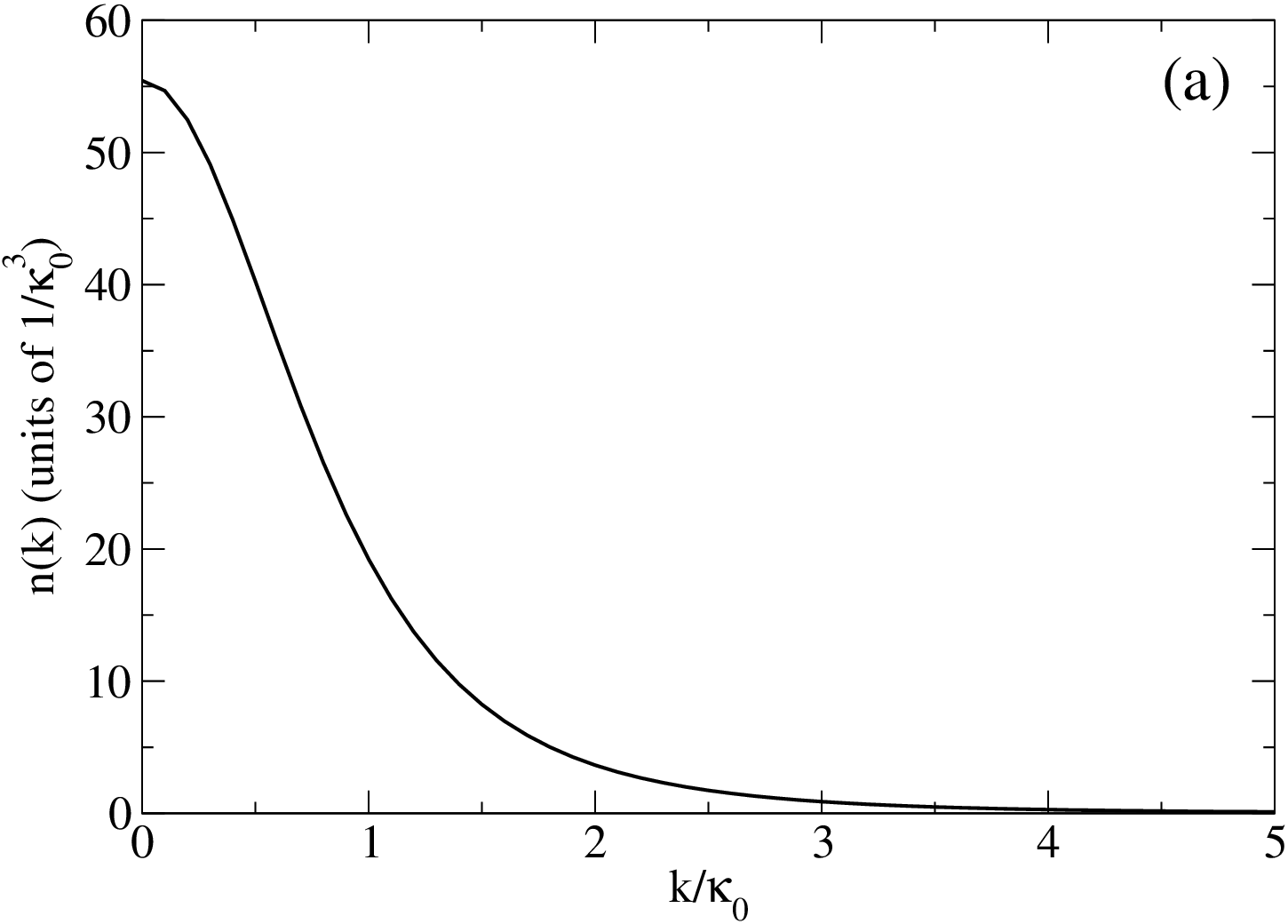}
\hfill
\includegraphics[width=0.46\linewidth,clip=]{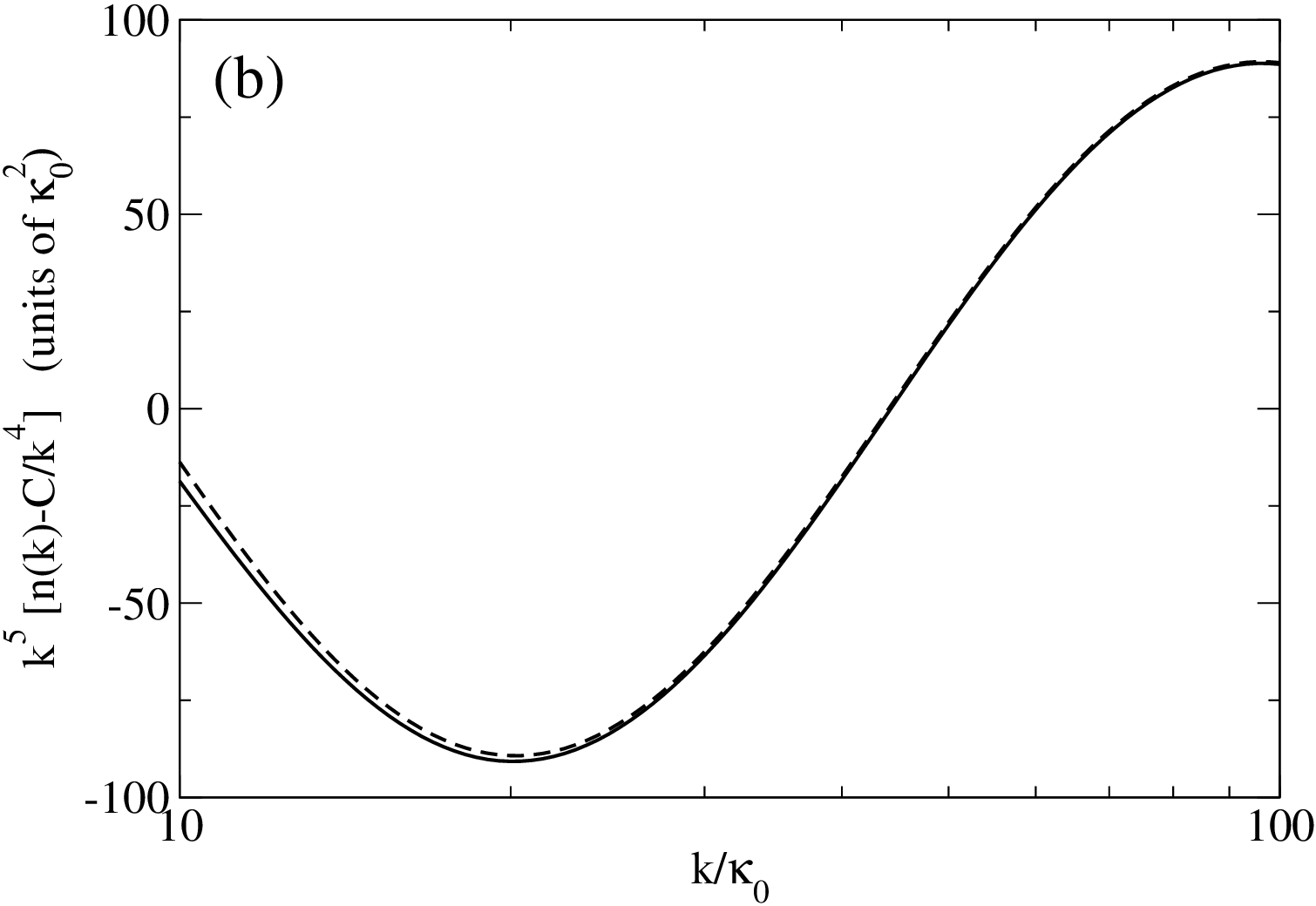}
\caption{For a free space Efimov trimer at rest
composed of three bosonic particles of mass $m$
interacting {\sl via} a zero-range, 
infinite scattering length potential, single-particle momentum distribution $n(\kk)$ 
as a function of $k$. (a) Numerical calculation from the expression for
$n(\kk)$ appearing in Eq.(\ref{eq:decomp}).
(b) Numerical calculation
(solid line) and asymptotic behavior (\ref{eq:nk_efi}) (dashed line),
with the horizontal axis in log scale.
The unit of momentum is $\kappa_0$, such that the trimer energy
is $-\hbar^2\kappa_0^2/m$.
\label{fig:nk}}
\end{figure*}

The integral expression of $n(\kk)$ derived in section \ref{sec:nk}
allows a straightforward and very precise
numerical calculation of the single-particle momentum distribution 
for an infinite scattering length Efimov trimer, once all the doable 
angular integrations have been performed in spherical coordinates of
polar axis $\kk_1$. The result is shown for
low values of $k$ in Fig.\ref{fig:nk}a, and for high values of $k$ 
in Fig.\ref{fig:nk}b. In particular,  Fig.\ref{fig:nk}b was constructed
to show how $n(\kk)$ approaches the asymptotic
behavior (\ref{eq:nk_efi}) derived in the next subsection, that is to reveal
the existence of a $1/k^5$ sub-leading oscillating term. Note that
the accuracy of the numerics may be tested from (\ref{eq:n_ori}) 
and from the explicit analytical expressions of $n_{\rm I}$
(given in (\ref{eq:nI_expl})), of $n_{\rm II}$ and of $n_{\rm III}$ (not given).

\subsection{Large momentum behavior of $n(\kk)$} 

Starting from the integral representation Eqs.(\ref{eq:nI},\ref{eq:nII},\ref{eq:nIII},\ref{eq:nIV}),
we show in the Appendix \ref{app:nk_high} 
that the single-particle momentum distribution has the asymptotic expansion at large wavevectors:
\be
n(\kk)\underset{k\to\infty}{=} \frac{C}{k^4}+
\frac{D}{k^5} \cos\left[2|s_0|\ln(\sqrt{3}k/\kappa_0)+\varphi\right]+
\ldots
\label{eq:nk_efi}
\ee
where we recall that the trimer energy is $E_{\rm trim}=-\hbar^2 \kappa_0^2/m$,
and the quantities $C$, $D$ and $\varphi$ derived in the Appendix \ref{app:nk_high}
are given by
\bea
C/\kappa_0&=&
8\pi^2 \sinh(|s_0|\pi/2) \tanh(|s_0|\pi)/\Big[\cosh(\frac{|s_0|\pi}{2}) \nonumber \\
\label{eq:Cexact}
&& +\frac{|s_0|\pi}{2} \sinh(\frac{|s_0|\pi}{2}) - \frac{4\pi}{3\sqrt{3}}
\cosh(\frac{|s_0|\pi}{6})\Big] \\
&=&
53.09722846003081\ldots
\\
D/\kappa_0^2 &\simeq& -89.26260
\\
\varphi &\simeq& -0.8727976.
\eea
The crucial point is that $D\neq 0$: Due to the Efimov effect,
the momentum distribution has a slowly decaying $O(1/k^5)$ oscillatory subleading tail. 

\subsection{Breakdown of the usual energy-momentum distribution relation}

In \cite{Leyronas} it was proposed that the expression of the energy as a functional
of the momentum distribution, derived in \cite{Tan2} for equal mass spin 1/2 fermions,
also holds for bosons (apart from the appropriate change of numerical factors). 
In the present case of a free space Efimov trimer at rest with an infinite scattering length,
the energy formula of \cite{Leyronas} reduces to
\be
\label{eq:leyronas}
E_{\rm trim} 
\stackrel{?}{=} \int \frac{d^3k}{(2\pi)^3} \frac{\hbar^2 k^2}{2m} \Big[n(\kk)-\frac{C}{k^4}\Big].
\ee
We have however put a question mark, because the asymptotic
expansion (\ref{eq:nk_efi}) implies that the integral in (\ref{eq:leyronas}) is not well-defined: 
After the change of variables $x=\ln(\sqrt{3}k/\kappa_0)$, the integrand behaves for $x\to+\infty$ 
as a linear superposition of $e^{2i |s_0| x}$ and $e^{-2i |s_0| x}$, that is 
as a periodic function of $x$ oscillating around zero.  This was overlooked in~\cite{Leyronas}.

At first sight, however, this does not look too serious: One often argues, when one faces
the integral of such an oscillating function of zero mean, that the oscillations 
at infinity simply average to zero.
More precisely, let us define the cut-off dependent energy functional
\be
E(\Lambda) = \int_{k<\Lambda} \frac{d^3\!k}{(2\pi)^3}  \frac{\hbar^2 k^2}{2m}
\left[n(\kk)-\frac{C}{k^4}\right],
\label{eq:elam}
\ee
where the integration is limited to wavevectors $\kk$ of modulus less than the cut-off.
For $\Lambda\to+\infty$, $E(\Lambda)$ is asymptotically an oscillating function of the 
logarithm of $\Lambda$, oscillating around a mean value $\bar{E}$.
The naive expectation would be that the trimer energy $E_{\rm trim}$ equals $\bar{E}$.
This naive expectation is equivalent to the usual trick used to regularize oscillating integrals, 
consisting here in introducing a convergence factor $e^{-\eta \ln(\sqrt{3}k/\kappa_0)}$ in the integral
without momentum cut-off and then taking the limit $\eta\to 0^+$:
\be
\lim_{\eta\to 0^+} \int_{\mathbb{R}^3} \frac{d^3\!k}{(2\pi)^3}  \frac{\hbar^2 k^2}{2m}
\left[n(\kk)-\frac{C}{k^4}\right] e^{-\eta \ln(\sqrt{3}k/\kappa_0)} = \bar{E}.
\label{eq:espoir}
\ee

\begin{figure}[htb]
\includegraphics[width=0.8\linewidth,clip=]{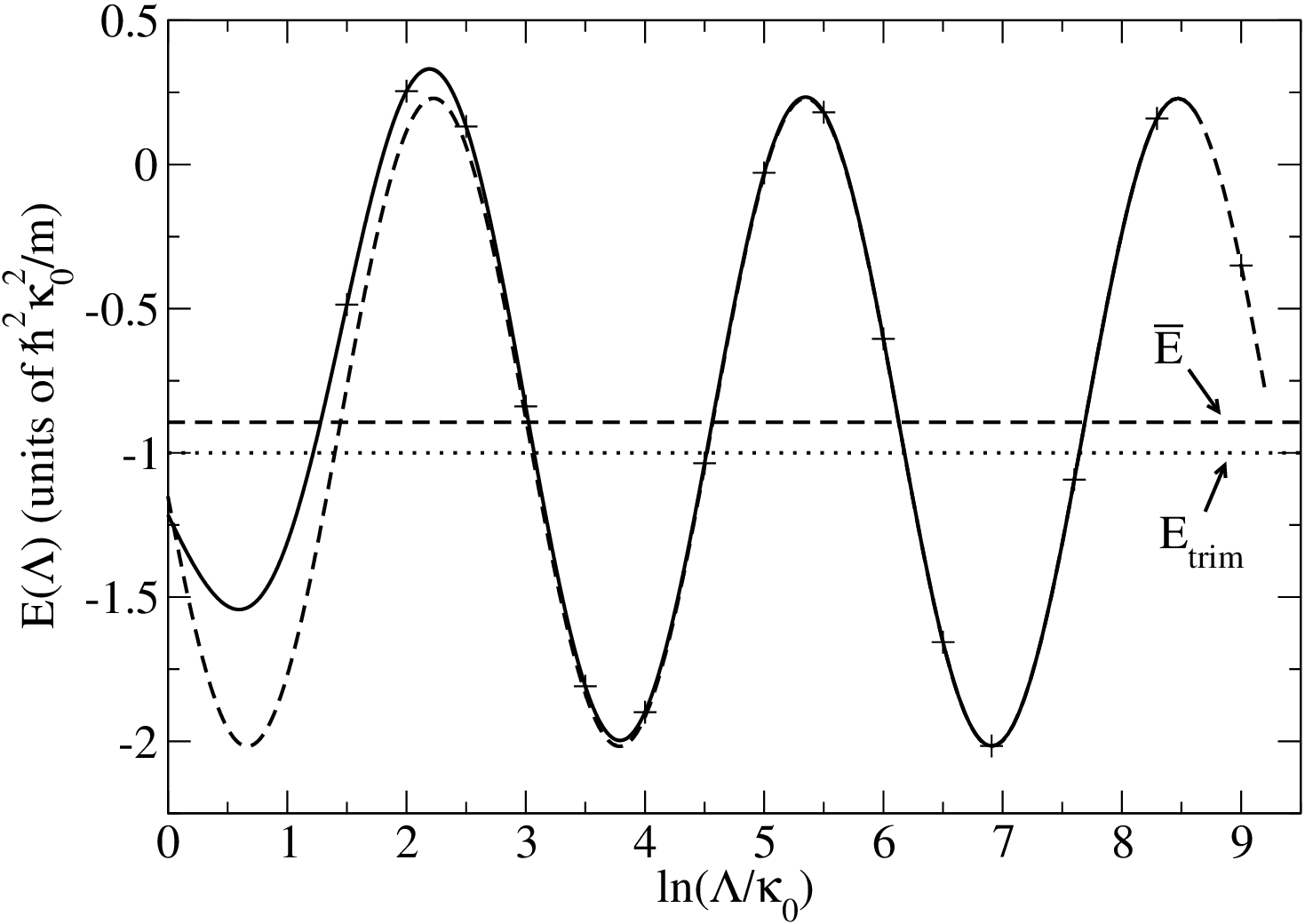}
\caption{Cut-off dependent energy $E(\Lambda)$ as defined in (\ref{eq:elam}) for a free space infinite scattering length Efimov trimer
with a zero-range interaction, as a function of the logarithm of the momentum cut-off $\Lambda$.
Solid line: numerical result obtained {\sl via} a calculation of the momentum distribution
$n(\kk)$. Symbols $+$: direct numerical calculation of $E(\Lambda)$ as exposed
in the Appendix \ref{app:brutale}. Dashed sinusoidal line: asymptotic oscillatory behavior of $E(\Lambda)$ 
for large $\Lambda$, obtained by omitting $O(1/\Lambda)$ in (\ref{eq:elamasympt}). 
Dashed horizontal line: mean value $\bar{E}$ around which $E(\Lambda)$ oscillates
at large $\Lambda$. The values of $\bar{E}$ obtained analytically 
(\ref{eq:ebar_analy}) and numerically (\ref{eq:nest}) are indistinguishable at the scale
of the figure, and clearly deviate from the dotted line giving the true energy
$E_{\rm trim}$
of the trimer, exemplifying the failure of a at a first sight convincing application
of an energy formula for bosons in three dimensions. 
The unit of momentum $\kappa_0$ is such that the true trimer energy
is $E_{\rm trim}=-\hbar^2 \kappa_0^2/m$.
\label{fig:elam}}
\end{figure}

To test the naive regularization procedure, we performed a numerical calculation of $E(\Lambda)$, 
using the result (\ref{eq:decomp}) to perform a very accurate numerical calculation of $n(\kk)$. 
The result is shown as a solid line in Fig.\ref{fig:elam}.
We also developed a more direct technique allowing a numerical calculation
of $E(\Lambda)$ without the knowledge of $n(\kk)$, see Appendix \ref{app:brutale}:
The corresponding results are represented as $+$ symbols in Fig.\ref{fig:elam}
and are in perfect agreement with the solid line.
As expected, $E(\Lambda)$ is asymptotically an oscillating function of the logarithm of $\Lambda$,
oscillating around a mean value $\bar{E}$.

To formalize,  we introduce an arbitrary, non-zero value
$k_{\rm min}$ of the momentum. We define $\delta n(k)= n(\kk)-C/k^4$ for $k<k_{\rm min}$,
and for $k>k_{\rm min}$:
\be
\delta n(k)=n(\kk) - \left\{\frac{C}{k^4}+ \frac{D}{k^5} \cos\left[2|s_0|\ln(\sqrt{3}k/\kappa_0)+
\varphi\right]\right\}.
\label{eq:defdn}
\ee
The introduction of $k_{\rm min}$ ensures that the integral of $k^2 \delta n(k)$ over
all $\kk$ converges around
$\kk=\mathbf{0}$. The subtraction of the asymptotic behavior of $n(\kk)$ up to order $O(1/k^5)$
for $k>k_{\rm min}$ ensures that the integral of $k^2 \delta n(k)$ 
over all $\kk$ converges at infinity.
As a consequence we get in the large cut-off limit
\be
E(\Lambda) = \bar{E} + \frac{\hbar^2 D}{8\pi^2 m |s_0|} 
\sin[2|s_0| \ln(\sqrt{3}\Lambda/\kappa_0)+\varphi] + O(1/\Lambda),
\label{eq:elamasympt}
\ee
with
\begin{multline}
\bar{E} = -\frac{\hbar^2 D}{8\pi^2 m |s_0|} \sin[2|s_0| \ln(\sqrt{3}k_{\rm min}/\kappa_0)+\varphi]
\\
+ \int_0^{+\infty} dk\,
\frac{\hbar^2 k^4}{4\pi^2 m}\delta n(k).
\label{eq:qveb}
\end{multline}
From this last equation (\ref{eq:qveb}) and the numerical calculations of $n(\kk)$ 
first up to $k=1000\kappa_0$ and
then up to $k\simeq 5500\kappa_0$, 
we get two slightly different values of $\bar{E}$, which gives
an estimate with an error bar 
\cite{astuce}:
\be
\bar{E} \simeq 0.89397(3) E_{\rm trim}.
\label{eq:nest}
\ee
The conclusion is that $\bar{E}$ (significantly) differs 
from $E_{\rm trim}$:
The naive regularization of the energy formula proposed
in \cite{Leyronas} does not give the correct value of the trimer energy.

An analytical representation of $\bar{E}$  in terms
of single integrals can be obtained, see Appendix \ref{app:analy}.
This analytical calculation gives a physical explanation of the failure of the 
naive regularization: It is inconsistent 
to add by hand the regularization factor 
$e^{-\eta \ln(\sqrt{3}k/\kappa_0)}$ at the last stage, that is in the integrand of
(\ref{eq:leyronas}). To be consistent, the momentum cut-off function has to be introduced
at the level of the three-body problem.
Then the subleading $1/k^5$ term in the momentum distribution
acquires a small non-oscillating component, of order $\eta/k^5$,  that
gives a non-zero contribution to the integral (\ref{eq:leyronas}) for $\eta\to 0^+$,
since $\int_{k_{\rm min}}^{+\infty} dk\, k^4 \frac{\eta}{k^5} e^{-\eta \ln(\sqrt{3}k/\kappa_0)}$
does not tend to zero in this limit.
The resulting integral representation of $\bar{E}$ confirms the numerical result
and allows to evaluate $\bar{E}$ with a better precision:
\begin{widetext}
\bea
\bar{E}/ E_{\rm trim}  &=& 
1 - \frac{4\cosh(|s_0|\pi/2)}{\pi\sqrt{3}} 
\Big[\cosh(|s_0|\pi/2)+\frac{|s_0|\pi}{2}\sinh(|s_0|\pi/2)-\frac{4\pi}{3\sqrt{3}}\cosh(|s_0|\pi/6)\Big]^{-1} \nonumber \\
&\times& \int_0^{+\infty} dq\, \left\{  \frac{2\ln(q)\cos(|s_0|\ln q)}{1+q^2+q^4} 
+\frac{16 q \ln[(1+q^2)/4]}{|s_0|(1+q^2)(q^2+3)^2}\sin\left(|s_0|\ln\frac{1+q}{|1-q|}\right)
\right.
\nonumber \\
\label{eq:ebar_analy}
&& \left. + 4 \ln\left( \frac{1+q^2}{|1-q^2|}\right) \cos\left(|s_0|\ln\frac{1+q}{|1-q|}\right)
\left[\frac{1}{2(q^2+3)}+\frac{\ln[2(1+q^2)/(q^2+3)]}{1-q^2}\right]\right\} \\
&=& 0.8939667780883\ldots
\eea
\end{widetext}

\section{Conclusion}
\label{sec:conclusion}

We have calculated the single-particle momentum distribution $n(\kk)$ 
for the free space Efimov trimer states
of same spin state bosons interacting {\sl via} a zero-range potential with an infinite scattering length.
The asymptotic behavior of $n(\kk)$ at large wavevectors, that we determined
with good precision, is of particular interest: In addition to the $C/k^4$ 
tail expected from two-body physics, 
it has a subleading oscillating $1/k^5$ contribution,  which is a signature
of Efimov physics that one may try to observe experimentally. 

We obtained the analytical expression for the coefficient $C$, see (\ref{eq:Cexact}).
This coefficient can also be obtained by a direct calculation in
position space, using the fact that $C$ is proportional to $\int d^3\rho\, |B(\rho)|^2$
\cite{WernerThese,tangen}.
This result allows to calculate the trimer energy for a finite scattering length $a$, to first order in $1/a$,
thanks to the relation \cite{tangen}
\be
\left(\frac{\partial E_{\rm trim}}{\partial (-1/a)}\right)_{R_t} = 
\frac{\hbar^2C}{8\pi m}
\ee
where the derivative is taken for a fixed value of the three-body parameter $R_t$.
In other words, we obtained analytically the derivative at $-\pi/2$ of Efimov's
universal function $\Delta(\xi)$ \cite{BH}. In numerical form, it gives
$\Delta'(-\pi/2)=2.125850069373\ldots$, which refines the previously known numerical estimate 
$\simeq 2.12$ \cite{BH}.
Furthermore, the existence of the $1/k^5$ subleading term leads to a failure of the
relation proposed in \cite{Leyronas} expressing the energy as a functional of the single-particle
momentum distribution, which was not obvious {\sl a priori}.

We have considered here the particular case of Efimov trimers. 
The coefficient of the $1/k^5$ subleading term was however obtained
in Appendix \ref{app:nk_high} by taking the zero-energy limit $\kappa_0\to 0$.
We thus expect that the phenomenology of the $1/k^5$ subleading term persists, not only for any other
three-body system subjected to Efimov physics (such as two fermions of mass $M$ and a lighter
particle of mass $m$ with a mass ratio $M/m>13.607$ \cite{Efimov2,PetrovEfim}), 
but also for a macroscopic Bose gas, 
at least for strong enough interactions \cite{Cornell_Bose,Chevy} so as to make the subleading
$1/k^5$ term sizeable and maybe accessible to measurements.
More precisely, we conjectured in \cite{tangen} that there is a subleading oscillating
$1/k^5$ term in the tail of the momentum distribution [as in Eq.(\ref{eq:nk_efi})] for any $N$-body quantum state subjected to
the Efimov effect, that is obeying three-body contact conditions involving the three-body parameter $R_t$, with
a coefficient $D$ related to $\partial E/\partial(\ln R_t)$ through a simple proportionality factor. Moreover, this derivative was
related in \cite{tangen} to a three-body analog of the contact, defined from the prefactor appearing in the three-body
contact condition.

\noindent {\bf Note:} After submission of this work, (i) the expression of the coefficient of the $1/k^5$
subleading tail of $n(\kk)$ in terms of the derivative of the energy with respect
to the logarithm of the three-body parameter $R_t$, and (ii) the appropriate energy formula taking into
account the Efimovian subleading $1/k^5$ term, appeared in \cite{platter_efim} for a
Bose gas with an arbitrary number of particles.

\acknowledgments
{We thank S. Tan for useful discussions. F.W. is supported by NSF under Grant No. PHY-0653183. 
Y.C. is a member of IFRAF and acknowledges support from ERC Project FERLODIM N.228177.}

\appendix

\section{Leading and next-to-leading terms for $n(\kk)$ at large momentum}
\label{app:nk_high}

Here we derive the asymptotic expansion (\ref{eq:nk_efi}).
We shall take the large $k_1$ limit, or equivalently formally the $\kappa_0\to 0$ limit
for a fixed $k_1$. From the asymptotic behavior (\ref{eq:agk})  we see that $\tilde{B}^2(k_1)$
involves a sum of ``oscillating" terms involving $k_1^{2s_0}$ or $k_1^{-2 s_0}$, and of
``non-oscillating" terms. We shall calculate first the resulting non-oscillating contribution, then
the resulting oscillating one, up to order $1/k_1^5$ included.

\subsubsection{Non-oscillating contribution up to $O(1/k_1^5)$}

We consider the small $\kappa_0$ limit successively for each of the four components
of $n(k_1)$ in (\ref{eq:decomp}).

\noindent{\bf Contribution I:} Taking directly  $\kappa_0\to 0$ in the integral defining $n_{\rm I}$, replacing
$\tilde{B}(k_1)$ by its asymptotic behavior (\ref{eq:agk}) and averaging out the oscillating
terms $k_1^{\pm 2 s_0}$ gives the leading behavior
\be
\langle n_{\rm I}(\kk_1)\rangle \simeq \frac{3\sqrt{3}}{8\pi} |\mathcal{N}_\psi|^2 \frac{4\pi^5}{k_1^5}.
\ee

\noindent{\bf Contribution II:}
In the integrand of (\ref{eq:nII}), we use the splitting
\begin{multline}
(k_1^2+k_2^2+\kk_1\cdot\kk_2+\kappa_0^2)^{-2}=k_1^{-4}  \\ + 
\left[(k_1^2+k_2^2+\kk_1\cdot\kk_2+\kappa_0^2)^{-2}-k_1^{-4}\right].
\end{multline}
The first term in the right hand side gives a contribution exactly scaling as $1/k_1^4$.
In the contribution of the second term in the right hand side, one may take the limit
$\kappa_0\to 0$ and replace $\tilde{B}^2(\sqrt{3} k_2/2)$ by its asymptotic expression to get
the subleading $1/k_1^5$ contribution. Performing the change of variable $\kk_2 = k_1 \qq$
in the integral and averaging out the oscillating terms
$k_1^{\pm 2 s_0}$ gives
\be
\langle n_{\rm II}(k_1) \rangle = \frac{C}{k_1^4} 
-\frac{3\sqrt{3}}{2\pi} |\mathcal{N}_\psi|^2 \frac{4\pi^5}{k_1^5} + o(1/k_1^5),
\ee
with
\be
C = 3 (\sqrt{3}/2)^6 \int\frac{d^3 k_2}{(2\pi)^3} 2\tilde{B}^2(\sqrt{3}k_2/2).
\label{eq:Cint}
\ee
We calculate $C$ from the exact expression (\ref{eq:Bt}) of $\tilde{B}$: We integrate over solid
angles and we use the change of variables $\frac{\sqrt{3}}{2} k_2 = \kappa_0 \sinh\alpha$,
where $\alpha$ varies from zero to $+\infty$, to take advantage of the fact that
\be
\tilde{B}(\kappa_0\sinh\alpha) = -\frac{2\pi^{5/2}\mathcal{N}_\psi}{\kappa_0^2 \sinh\alpha\cosh\alpha}
\left(e^{s_0\alpha}-e^{-s_0\alpha}\right).
\label{eq:sacdv}
\ee
This leads to
\be
C =  12\pi^3 (\sqrt{3}/2)^3 \frac{|\mathcal{N}_\psi|^2}{\kappa_0} 
\int_0^{+\infty} d\alpha\, \frac{2-(e^{2s_0\alpha}+\mbox{c.c.})}{\cosh\alpha},
\ee
where we used the fact that $\mathcal{N}_\psi^2=-|\mathcal{N}_\psi|^2$.
The resulting integral over $\alpha$ may be extended over the whole real axis because
the integrand is an even function of $\alpha$; it may then be evaluated by using the general result
(that we obtained with contour integration)
\be
\label{eq:Utile}
K(\theta,s) \equiv \int_{-\infty}^{+\infty} d\alpha\, \frac{e^{is\alpha}}{\cosh\alpha +\cos\theta}
= \frac{2\pi}{\sin\theta} \frac{\sinh(s\theta)}{\sinh(s\pi)}
\ee
where $s$ is a real number and $\theta\in]0,\pi[$. One simply has to take $\theta=\pi/2$,
$s\to 0$ and $s=2|s_0|$ respectively. We get
\be
\label{eq:Cprov}
C = \frac{24 \pi^4}{\kappa_0}  \left(\frac{\sqrt{3}}{2}\right)^3 
\frac{\sinh^2(|s_0|\pi/2)}{\cosh(|s_0|\pi)}\,
|\mathcal{N}_\psi|^2.
\ee
This, together with (\ref{eq:norma}), leads to the explicit expression (\ref{eq:Cexact}) for $C$.

\noindent{\bf Contribution III:}
We directly take the limit $\kappa_0\to 0$ and we replace the factors $\tilde{B}$ by
their asymptotic expressions in (\ref{eq:nIII}). After the change of variable
$\kk_2=k_1 \qq$, angular integration and averaging  out of the oscillating
terms $k_1^{\pm 2 s_0}$, this gives
\be
\langle n_{\rm III}(\kk_1)\rangle =
\frac{9}{2\pi^2} \frac{4\pi^5|\mathcal{N}_\psi|^2}{k_1^5} \int_0^{+\infty} dq \, \frac{q^{s_0}+q^{-s_0}}{q^4+q^2+1}
+o(1/k_1^5).
\ee
In this result, we change the integration variable setting $q=e^\alpha$, where $\alpha$ varies from $-\infty$ to
$+\infty$. The odd component of the integrand (involving $\sinh\alpha$) gives a vanishing contribution.
The even component of the integrand involves a rational fraction of $\cosh\alpha$ to which we apply
a partial fraction decomposition. Then we use (\ref{eq:Utile}) to obtain
\begin{multline}
n_{\rm III}(k_1) = 
\frac{4\pi^5 |\mathcal{N}_\psi|^2}{k_1^5} \frac{3\sqrt{3}}{2\pi}
\frac{\sinh(\pi|s_0|/3)+\sinh(2\pi|s_0|/3)}{\sinh(\pi|s_0|)}  \\
+ o(1/k_1^5).
\end{multline}

\noindent{\bf Contribution IV:}
We directly take the limit $\kappa_0\to 0$ and we replace the factors $\tilde{B}$ by
their asymptotic expressions in (\ref{eq:nIV}). We perform the change of variable
$\kk_2=(k_1/2)\qq$, we average out the oscillating terms $k_1^{\pm 2 s_0}$.
The angular integration in spherical coordinates of axis the direction of $\kk_1$
may be performed using
\be
\int dv\, \left(\frac{1+v}{1-v}\right)^{s_0/2}(1-v^2)^{-1} = \left(\frac{1+v}{1-v}\right)^{s_0/2}/s_0,
\ee
where the variable $v$ is restricted to the interval $(-1,1)$. This leads to
\begin{multline}
\langle n_{\rm IV}(\kk_1)\rangle =
\frac{4\pi^5 |\mathcal{N}_\psi|^2}{k_1^5}
\frac{36}{\pi^2}
\int_0^{+\infty} dq\, \frac{q}{q^2+1} (q^2+3)^{-2} \\
\times \left[s_0^{-1} \left(\frac{q+1}{|q-1|}\right)^{s_0}+
\mbox{c.c.}\right]
+o(1/k_1^5).
\end{multline}
Calculating this integral directly is not straightforward because of the occurrence of the absolute
value $|q-1|$. We thus split the integration domain in two intervals.
For $q\in [0,1]$ we set $q=(X-1)/(X+1)$ (an increasing function of $X$,
where $X$ spans $[1,+\infty]$).
For $q\in [1,+\infty]$ we set $q=(X+1)/(X-1)$ (a decreasing function of $X$,
where $X$ here also spans $[1,+\infty]$).
Then
\begin{multline}
\langle n_{\rm IV}(\kk_1)\rangle=
\frac{4\pi^5 |\mathcal{N}_\psi|^2}{k_1^5}
\frac{9}{2\pi^2}
\int_1^{+\infty} \frac{dX}{X}\, \Big\{ (X-X^{-1})\\ \times  \frac{(X^2-1+X^{-2})}{(X^2+1+X^{-2})^2} 
\left[s_0^{-1} X^{s_0} - s_0^{-1} X^{-s_0}\right]\Big\} + o(1/k_1^5).
\end{multline}
We then set $X=e^\alpha$, where $\alpha$ ranges from zero to $+\infty$,
and we use the fact that the resulting integrand is an even function of $\alpha$
to extend the integral over the whole real axis.
We integrate by parts, integrating the factor $\sin(\alpha|s_0|)$,
and we perform a partial fraction decomposition of the resulting rational fraction of
$\cosh\alpha$.
Using (\ref{eq:Utile}) and its derivatives with respect to $\theta$, we get
\begin{multline}
\langle n_{\rm IV}(\kk_1)\rangle =
\frac{72\pi^5|\mathcal{N}_\psi|^2}{k_1^5}
\Big\{\cosh(\frac{2\pi|s_0|}{3})-\cosh(\frac{\pi|s_0|}{3}) \\
-\frac{\sqrt{3}}{6} |s_0| \Big[\sinh(\frac{2\pi|s_0|}{3})+\sinh(\frac{\pi|s_0|}{3})\Big]\Big\}/
[2\pi|s_0|\sinh(\pi|s_0|)] \\
+ o(1/k_1^5).
\end{multline}

\noindent{\bf Sum of the four contributions:}
Summing up the non-oscillating terms in $1/k_1^5$ of the contributions 
$n_{\rm I}$, $n_{\rm II}$, $n_{\rm III}$
and $n_{\rm IV}$, we obtain as a global prefactor
\be
\mathcal{S} = -\frac{\sqrt{3}}{8} +
\frac{\cosh(2\pi|s_0|/3)-\cosh(\pi|s_0|/3)}{|s_0|\sinh(\pi|s_0|)}.
\ee
Multiplying (\ref{eq:def_ms0}) on both sides by $\sinh(|s_0|\pi/2)$ 
and using 
\be
2\sinh a\sinh b = \cosh(a+b)-\cosh(a-b), \ \ \ \forall a,b
\ee
we find that $\mathcal{S}$ is exactly zero. As a consequence,
the non-oscillating part of the momentum distribution of an infinite
scattering length Efimov trimer behaves at large $k_1$ as
\be
\langle n(\kk_1)\rangle = \frac{C}{k_1^4} + o(1/k_1^5).
\ee

\subsubsection{Oscillating contribution at large $k_1$}

In the large $k_1$ tail of the momentum distribution, we now include
{\sl oscillating} terms, having oscillating factors such as $k_1^{\pm 2 s_0}$.
The calculation techniques are the same as in the previous subsection, so that
we give here directly the result.
We find that the leading oscillating terms scale as $1/k_1^5$:
\begin{multline}
n(\kk_1)-\langle n(\kk_1)\rangle 
=
-\frac{12\pi^5}{k_1^5} |\mathcal{N}_\psi|^2 
\left[\mathcal{A}\left(\frac{k_1 \sqrt{3}}{\kappa_0}\right)^{2s_0}+\mbox{c.c.}
\right] \\
+o(1/k_1^5)
\end{multline}
where the complex amplitude $\mathcal{A}$ is the sum of the contributions
coming from each of the four components (\ref{eq:nI},\ref{eq:nII},\ref{eq:nIII},\ref{eq:nIV})
of the moment distribution,
\be
\mathcal{A} = \mathcal{A}_{\rm I} + \mathcal{A}_{\rm II} + 
\mathcal{A}_{\rm III} + \mathcal{A}_{\rm IV}.
\ee
We successively find
\bea
\mathcal{A}_{\rm I} &=&  
\frac{3}{8\pi^2} \int_0^{+\infty} \!\!\!\! dq\,
\frac{q^2}{q^4+q^2+1} = \frac{\sqrt{3}}{16\pi}, \\
\mathcal{A}_{\rm II} &=& 
\frac{3}{4\pi^2} \int_0^{+\infty}\!\!\!\!  dq\,
\frac{q^{2s_0}}{q^2} \left[(q^4+q^2+1)^{-1}-1\right]  \\
&=& -\frac{\sqrt{3}}{4\pi} \frac{\sinh(4\pi|s_0|/3)+\sinh(2\pi|s_0|/3)}{\sinh(2\pi|s_0|)}, \\
\mathcal{A}_{\rm III} &=& 
\frac{3}{2\pi^2} \int_0^{+\infty}\!\!\!\!  dq\,
\frac{q^{s_0}}{q^4+q^2+1} 
= \frac{\sqrt{3}}{4\pi\sinh(\pi |s_0|)} \nonumber \\
&\times& \Big\{\sinh(\frac{2\pi|s_0|}{3})+\sinh(\frac{\pi|s_0|}{3}) \nonumber \\
&& -i\sqrt{3} \Big[\cosh(\frac{2\pi|s_0|}{3})-\cosh(\frac{\pi|s_0|}{3})\Big]\Big\}, \\
\label{eq:dur}
\mathcal{A}_{\rm IV} &=&
\frac{12}{\pi^2}\ 2^{-2 s_0}
\int_0^{+\infty} \!\! dq\, \Big[ \frac{q (1+q^2)^{s_0}}{(q^2+3)^2(q^2+1)} \nonumber  \\
&& \times \int_0^{2q/(1+q^2)}\!\!\!\!   dv\, \frac{(1-v^2)^{s_0/2}}{1-v^2}\Big] 
\label{eq:over_v}\\
&\simeq&  0.0243657158 - 0.0698680251 i.
\eea
We have calculated analytically all these
integrals, except for (\ref{eq:dur}) where the angular integration gives rise to the integral over $v$ and thus to a
difficult hypergeometric function.
Finally
\be
\mathcal{A} \simeq 0.1022397786 - 0.1218775240 i.   
\ee

\section{Direct calculation of $E(\Lambda)$}
\label{app:brutale}

To calculate the cut-off dependent energy $E(\Lambda)$ defined in (\ref{eq:elam})
for an infinite scattering length Efimov trimer, the method consisting
in calculating the momentum distribution $n(\kk)$ and then 
integrating (\ref{eq:elam}) is numerically demanding: A double integral
has to be performed to obtain $n(\kk)$, see (\ref{eq:nIV}), so that
the evaluation of $E(\Lambda)$ results in a triple integral.
A more direct formulation, involving only a double integration, is proposed here.
One simply rewrites (\ref{eq:elam}) as
\be
E(\Lambda) = \int_{\mathbb{R}^3} \frac{d^3\!k}{(2\pi)^3}  f(k)\frac{\hbar^2 k^2}{2m}
\left[n(\kk)-\frac{C}{k^4}\right]
\label{eq:wcof}
\ee
where the function $f(k)$ is equal to unity for $0\leq k\leq \Lambda$
and is equal to zero otherwise.
Then one plugs in (\ref{eq:wcof}) the expression (\ref{eq:decomp}) of $n(\kk)$, also
replacing $C$ with its integral expression (\ref{eq:Cint}). An integration over two
vectors in $\mathbb{R}^3$ appears, so that
$E(\Lambda)=E_{\rm easy}(\Lambda) + E_{\rm hard}(\Lambda)$ with
\begin{multline}
E_{\rm easy}(\Lambda)= 
\frac{3^4}{4^3}
\int \frac{d^3k}{(2\pi)^3} f(k)\frac{\hbar^2 k^2}{2m}
\int \frac{d^3q}{(2\pi)^3}
\Big[-\frac{2\tilde{B}^2(\frac{\sqrt{3}}{2}q)}{k^4} \\
+ \frac{\tilde{B}^2(\frac{\sqrt{3}}{2}k) + 2 \tilde{B}^2(\frac{\sqrt{3}}{2}q)
+ 4 \tilde{B}(\frac{\sqrt{3}}{2}k) \tilde{B}(\frac{\sqrt{3}}{2}q)}{(k^2+q^2+
\kk\cdot\qq + \kappa_0^2)^2}\Big]
\end{multline} 
\begin{multline}
E_{\rm hard}(\Lambda)= 
\frac{3^4}{4^3}
\int \frac{d^3k}{(2\pi)^3} f(k)\frac{\hbar^2 k^2}{2m} 
\int \frac{d^3q}{(2\pi)^3} \\
\frac{2\tilde{B}(\frac{\sqrt{3}}{2}|\qq+\kk/2|) \tilde{B}(\frac{\sqrt{3}}{2}|\qq-\kk/2|)}
{(q^2+\frac{3}{4} k^2 +\kappa_0^2)^2}.
\label{eq:grosse_exp}
\end{multline}
The first part $E_{\rm easy}$ of this expression originates from the bits 
$n_{\rm I}$, $n_{\rm II}$,
$n_{\rm III}$ of the momentum distribution and from $C$; angular integrations
may be performed, one is left with a double integral over the moduli
$k$ and $q$. Taking $\kappa_0$ as a unit of momentum and $\hbar^2 \kappa_0^2/m$
as a unit of energy in what follows:
\begin{multline}
E_{\rm easy}(\Lambda) = \left(\frac{3}{4\pi}\right)^4
\int_0^{\Lambda} dk\, \frac{k^4}{2} \int_0^{+\infty} \!\!\!\! dq\, q^2 
\Big[
-\frac{2\tilde{B}^2(\frac{\sqrt{3}}{2}q)}{k^4} \\
+\frac{\tilde{B}^2(\frac{\sqrt{3}}{2}k) + 2 \tilde{B}^2(\frac{\sqrt{3}}{2}q)
+ 4 \tilde{B}(\frac{\sqrt{3}}{2}k) \tilde{B}(\frac{\sqrt{3}}{2}q)}{(k^2+q^2+1)^2-k^2q^2}
\Big]
\end{multline}
that we integrate numerically.
The second part $E_{\rm hard}(\Lambda)$ in (\ref{eq:grosse_exp})  
originates from the bit $n_{\rm IV}$ 
of the momentum distribution. Performing the change of variables
$\qq=(\kk_1-\kk_2)/2$ and $\kk=\kk_1+\kk_2$ ensures that the factors $\tilde{B}$
are now functions of the moduli $k_1$ and $k_2$ only, 
\begin{multline}
E_{\rm hard}(\Lambda) = \frac{3^4}{4^3} \int \frac{d^3k_1}{(2\pi)^3}
\int \frac{d^3k_2}{(2\pi)^3}
\frac{1}{2} (\kk_1+\kk_2)^2 f(|\kk_1+\kk_2|)  \\
\times \frac{2\tilde{B}(\frac{\sqrt{3}}{2}k_1)\tilde{B}(\frac{\sqrt{3}}{2}k_2)}
{(k_1^2+k_2^2+\kk_1\cdot\kk_2 + 1)^2}
\end{multline}
so that angular integrations
may again be performed, involving the integral
\bea
I(k_1,k_2) &=& \frac{1}{2} \int_{-1}^{1} du\, \frac{k_1^2+k_2^2+2k_1k_2 u}
{(k_1^2+k_2^2+k_1 k_2 u +1)^2}\nonumber \\
&\times&  f\left(\sqrt{k_1^2+k_2^2+2k_1k_2 u}\right) \\
&=&
\frac{1}{k_1 k_2}\,\Big[ \ln(1+k_1^2+k_2^2+k_1 k_2 u) \nonumber \\
&+& \frac{1+(k_1^2+k_2^2)/2}{1+k_1^2+k_2^2+k_1 k_2 u}
\Big]_{-1}^{\mathrm{max}[-1,\mathrm{min}(1,U)]}
\eea
where $u$ is the cosine of the angle between the vectors $\kk_1$ and $\kk_2$,
$U=[\Lambda^2-(k_1^2+k_2^2)]/(2 k_1 k_2)$, $\mathrm{max}(a,b)$ (resp. $\mathrm{min}(a,b)$)
is the largest (resp. smallest)
of the two numbers $a$ and $b$, 
and the notation $[F(u)]_a^b$
stands for $F(b)-F(a)$ for any function $F(u)$.
We also used the fact that $|\kk_1+\kk_2|\leq \Lambda$ if and only if
$u\leq U$.  This leads to
\begin{multline}
E_{\rm hard}(\Lambda) = \left(\frac{3}{4\pi}\right)^4
\int_0^{+\infty}\!\!\!\!\ dk_1 \, k_1^2 
\int_0^{+\infty}\!\!\!\!\ dk_2 \, k_2^2 \,
I(k_1,k_2) \\
\times \tilde{B}(\frac{\sqrt{3}}{2}k_1) \tilde{B}(\frac{\sqrt{3}}{2}k_2).
\end{multline}
Further simplifications may be performed. One can map the integration to the
domain $k_1\geq k_2$ since the integrand is a symmetric function of $k_1$ and
$k_2$. Then performing the change of variable $k_1=q+k/2$ and $k_2=q-k/2$,
and using the fact that $I(k_1,k_2)=0$ if $k_1-k_2>\Lambda$, we obtain the useful form
\begin{multline}
E_{\rm hard}(\Lambda) =  \left(\frac{3}{4\pi}\right)^4
2\int_0^{\Lambda}\!\!\!  dk\, \int_{k/2}^{+\infty} \!\!\! dq\,
(q^2-k^2/4)^2 \\ 
\times I(q+\frac{1}{2}k,q-\frac{1}{2}k) \tilde{B}\left[\frac{\sqrt{3}}{2}(q+k/2)\right]
\tilde{B}\left[\frac{\sqrt{3}}{2}(q-k/2)\right],
\end{multline}
that we integrate numerically.
A useful result to control the numerical error due to the truncation of the integral
over $q$ to a value $\gg \Lambda$ and $\gg 1$ is
$I(q+k/2,q-k/2) \underset{q\to +\infty}{\sim} (k^4-\Lambda^4)/(8 q^6)$.

\section{Analytical expression for $\bar{E}$} 
\label{app:analy}

As explained in the main text, the naive regularization (\ref{eq:espoir}) of the energy formula
gives an energy $\bar{E}$ that actually differs from the energy of the trimer $E_{\rm trim}$,
because the momentum space cut-off function was introduced at the last stage of the calculation.
Here we introduce a momentum cut-off function at the level of the three-body wavefunction,
simply by making the substitution 
\be
\tilde{B}(k) \to \tilde{B}_\eta(k)\equiv \tilde{B}(k)\,  e^{-\eta \alpha}
\ee
where we have set $k=\kappa_0 \sinh \alpha$ so that $\alpha = \ln [\sqrt{1+k^2/\kappa_0^2}+k/\kappa_0]$. 
For this {\sl consistent} regularization,
we expect that the usual energy formula holds in the limit of a vanishing $\eta$,
and this was checked explicitly in Appendix H of \cite{tangen}. This means that
\be
\lim_{\eta\to 0^+} E_\eta = E_{\rm trim} 
\ee
with
\be
E_\eta = \int_{\mathbb{R}^3} \frac{d^3k}{(2\pi)^3} \frac{\hbar^2 k^2}{2m} \Big[ n_\eta(\kk) - \frac{C_\eta}{k^4}\Big].
\ee
The single-particle momentum distribution $n_\eta$ is obtained by replacing the function $\tilde{B}$ by
$\tilde{B}_\eta$ in Eqs.(\ref{eq:decomp},\ref{eq:nI},\ref{eq:nII},\ref{eq:nIII},\ref{eq:nIV}). Its large
$k$ asymptotic behavior can be obtained along the lines of Appendix \ref{app:nk_high}:
$n_\eta(\kk) = n_\eta^{\rm asymp}(k) + O(1/k^6)$ with
\begin{multline}
n_\eta^{\rm asymp}(k) = \frac{C_\eta}{k^4}+
\frac{e^{-2\eta\ln(\sqrt{3}k/\kappa_0)}}{k^5} \Big\{\bar{D}_\eta\\  +
D_\eta \cos\left[2|s_0|\ln(\sqrt{3}k/\kappa_0)+\varphi_\eta\right]\Big\}.
\label{eq:nketaasymp}
\end{multline}
In the limit $\eta\to 0^+$ one has to recover (\ref{eq:nk_efi}) so that
the coefficient $\bar{D}_{\eta}$ of the non-oscillating $1/k^5$ contribution
tends to zero in that limit. However, as we shall see, $\bar{D}_{\eta}/\eta$ does not
tend to zero, which leads to the failure of the naive regularization.
The expressions of the other coefficients $C_\eta$, $D_\eta$ and $\varphi_\eta$ are not needed.

Following (\ref{eq:defdn}) we define $\delta n_\eta(k)$ as $n_\eta(\kk) -C_\eta/k^4$ for $k< k_{\rm min}$
and as $n_\eta(\kk) - n_\eta^{\rm asymp}(k)$ for $k> k_{\rm min}$.
This results in the splitting
\begin{multline}
E_\eta = \int_{\mathbb{R}^3} \frac{d^3k}{(2\pi)^3} \, \frac{\hbar^2 k^2}{2m} \delta n_\eta(k) \\
+ \frac{\hbar^2}{4\pi^2 m} \int_{x_{\rm min}}^{+\infty}
dx\, e^{-2\eta x} [\bar{D}_\eta + D_\eta \cos(2|s_0|x+\varphi_\eta)]
\label{eq:eta_exp}
\end{multline}
where the change of variable $x=\ln(\sqrt{3}k/\kappa_0)$ was used
so that $x_{\rm min}=\ln(\sqrt{3}k_{\rm min}/\kappa_0)$.
For $\eta\to 0^+$, we can replace in the right hand side of (\ref{eq:eta_exp})
$\delta n_\eta(k)$ with $\delta n(k)$ since
the first integral converges absolutely, but we cannot exchange 
the $\eta\to 0^+$ limit
and the integration in the second integral.
After explicit calculation of this second integral, we take
$\eta\to 0^+$ and  we recognize $\bar{E}$
from (\ref{eq:qveb}) so that
\be
E_{\rm trim} = \bar{E} + \frac{\hbar^2}{8\pi^2 m} \lim_{\eta\to0^+}
\frac{\bar{D}_\eta}{\eta}.
\label{eq:tec}
\ee

The last step is to calculate $\bar{D}_\eta$, with the same techniques as in the Appendix
\ref{app:nk_high}. One finds
\be
\bar{D}_\eta = \frac{9}{2\pi^2} |\mathcal{N}_\psi|^2 4\pi^5
\left[\frac{\pi}{4\sqrt{3}} + I_\eta + J_\eta + K_\eta\right]
\ee
with
\bea
I_\eta &=& \int_0^{+\infty} \!\!\!\! dq\, \frac{-(1+q^2)}{1+q^2+q^4} \, e^{-2\eta\ln q}
\\
J_\eta &=& \int_0^{+\infty} \!\!\!\! dq\, \frac{q^{s_0}+q^{-s_0}}{1+q^2+q^4} e^{-\eta\ln
q} \\
K_\eta &=& \int_0^{+\infty} \!\!\!\! dq\,
\frac{8q}{1+q^2} \frac{e^{-\eta \ln\left(\frac{1+q^2}{4}\right)}}{(q^2+3)^2}
\int_0^{\frac{2q}{1+q^2}} \!\! dv\, \frac{e^{-\eta\ln\sqrt{1-v^2}}}{1-v^2} \nonumber \\
&& \times \left[\left(\frac{1+v}{1-v}\right)^{s_0/2} + \mbox{c.c.}\right].
\label{eq:Keta}
\eea
The contributions $I_\eta$, $J_\eta$ and $K_\eta$ originate respectively
from the bits $n_{II}$, $n_{III}$ and $n_{IV}$ in the decomposition (\ref{eq:decomp})
generalized to $n_\eta$. Taking the derivative with respect to $\eta$
and then taking the limit $\eta\to 0^+$ gives (\ref{eq:ebar_analy}).
More precisely, one finds that 
\bea
\frac{dI_\eta}{d\eta}|_{\eta\to0^+}  & =& 0 \\
\frac{dJ_\eta}{d\eta}|_{\eta\to0^+}  &= & \int_0^{+\infty} dq\,
(-\ln q)\, \frac{q^{s_0}+q^{-s_0}}{1+q^2+q^4}
\eea
which, together with (\ref{eq:ebar_analy}) and (\ref{eq:tec}), suffices to determine
$\frac{dK_\eta}{d\eta}|_{\eta\to0^+}$ so that we do not reproduce here its lengthy expression.
The remarkable fact that $\frac{dK_\eta}{d\eta}|_{\eta\to0^+}$ may be expressed as
an integral over a single variable $q$, whereas the expression of $K_\eta$ for a general
$\eta$ in (\ref{eq:Keta}) involves a double integral, 
results from an integration by part over $q$ in (\ref{eq:Keta}),
taking the derivative of the bit $\int_0^{2q/(1+q^2)}dv\ldots$.


\begin{thebibliography}{0}
\expandafter\ifx\csname natexlab\endcsname\relax\def\natexlab#1{#1}\fi
\expandafter\ifx\csname bibnamefont\endcsname\relax
  \def\bibnamefont#1{#1}\fi
\expandafter\ifx\csname bibfnamefont\endcsname\relax
  \def\bibfnamefont#1{#1}\fi
\expandafter\ifx\csname citenamefont\endcsname\relax
  \def\citenamefont#1{#1}\fi
\expandafter\ifx\csname url\endcsname\relax
  \def\url#1{\texttt{#1}}\fi
\expandafter\ifx\csname urlprefix\endcsname\relax\def\urlprefix{URL }\fi
\providecommand{\bibinfo}[2]{#2}
\providecommand{\eprint}[2][]{\url{#2}}

\end{thebibliography}


\begin{thebibliography}{99}
\bibitem{Feshbach}
C. Chin, R. Grimm, P. Julienne, and E. Tiesinga,
Rev. Mod. Phys. {\bf 82}, 1225 (2010).

\bibitem{revueStringariFermions}
S. Giorgini, L. P. Pitaevskii, S. Stringari,
Rev. Mod. Phys. {\bf 80}, 1215 (2008).

\bibitem{Efimov_manips}
T. Kraemer, M. Mark, P. Waldburger, J. G. Danzl, C. Chin, B. Engeser, A. D. Lange, K. Pilch, A. Jaakkola, H.-C. Naegerl, R. Grimm,
Nature {\bf 440}, 315 (2006);
M. Zaccanti, B. Deissler, C. D'Errico, M. Fattori, M. Jona-Lasinio,
S. {M\"uller}, G. Roati, M. Inguscio, G. Modugno,
Nature Physics {\bf 5}, 586 (2009);
N. Gross, Z. Shotan,  S. Kokkelmans, L. Khaykovich, 
Phys. Rev. Lett.  {\bf 103}, 163202 (2009);
{\sl ibid.}, Phys. Rev. Lett. {\bf 105}, 103203 (2010); 
S. E. Pollack, D. Dries, R.G. Hulet, Science {\bf 326}, 1683 (2009).

\bibitem{Jochim}
T. Lompe,  T.  B. Ottenstein, F. Serwane, A. N. Wenz, G. Z\"urn, S. Jochim,
Science {\bf 330}, 940 (2010).

\bibitem{BH}
E. Braaten and H.-W. Hammer, Phys. Rept. {\bf 428}, 259 (2006).

\bibitem{Efimov1}
V. N. Efimov, Yad. Fiz. {\bf 12}, 1080 (1970) [Sov. J. Nucl. Phys. {\bf 12}, 
589 (1971)].

\bibitem{Efimov2}
V. Efimov, Nucl. Phys. A {\bf 210}, 157 (1973).

\bibitem{PetrovEfim}
D. Petrov, Phys. Rev. A {\bf 67}, 010703 (2003).

\bibitem{Jin_nk}
J. T. Stewart, J. P. Gaebler, T. E. Drake, D. S. Jin,
Phys. Rev. Lett. {\bf 104}, 235301 (2010).

\bibitem{Blume}
D. Blume and K. M. Daily, Phys. Rev. A {\bf 80}, 053626 (2009).

\bibitem{Esslinger}
I. Bloch, T.W. H\"ansch, T. Esslinger, Nature {\bf 403}, 166 (2000) ;
T. Donner, S. Ritter, T. Bourdel, A. \"Ottl, M. K\"ohl, T. Esslinger, 
Science {\bf 315}, 1556 (2007).

\bibitem{lien_g1}
In 1D, this contribution scales as $r^3$ for $r\to 0$ \cite{Olshanii}. 
In 3D, this contribution is linear in $r$ for $r\to 0$,
with a coefficient proportional to $C$ \cite{tangen,Tan2}.

\bibitem{Astra}
G. E. Astrakharchik, J. Boronat, J. Casulleras, and S. Giorgini, 
Phys. Rev. Lett. {\bf 95}, 230405 (2005).

\bibitem{contact_interaction}
S. Albeverio, F. Gesztesy, R. Hoegh-Krohn, and H. Holden, 
{\it Solvable models in quantum mechanics} (Springer-Verlag,
Berlin, 1988);
Y. Castin, in: Coherent atomic matter waves, Lecture Notes of 1999 
Les Houches Summer School, R. Kaiser, C.Westbrook,
and F. David eds. (EDP Sciences and Springer-Verlag, 2001);
Y. Castin, Comptes Rendus Physique {\bf 5}, 407 (2004);
L. Pricoupenko and Y. Castin, J. Phys. A {\bf 40}, 12863 (2007).

\bibitem{Olshanii}
M. Olshanii and V. Dunjko, Phys. Rev. Lett. {\bf 91}, 090401 (2003).

\bibitem{Tan1}
Shina Tan, Annals of Physics {\bf 323}, 2971 (2008).

\bibitem{Braaten_contact}
E. Braaten, L. Platter, Phys. Rev. Lett. {\bf 100}, 205301 (2008).

\bibitem{Tarruell}
F. Werner, L. Tarruell, and Y. Castin, Eur. Phys. J. B {\bf 68}, 401 (2009).

\bibitem{tangen}
F. Werner, Y. Castin, arXiv:1001.0774 (2010).

\bibitem{HammerBosons}
E. Braaten, H.-W. Hammer, and T. Mehen,
Phys. Rev. Lett. {\bf 88}, 040401 (2002).

\bibitem{Cornell_Bose}
S.B. Papp, J.M. Pino, R.J. Wild, S. Ronen, C.E. Wieman, D.S. Jin, and E.A. Cornell,
Phys. Rev. Lett. {\bf 101}, 135301 (2008).

\bibitem{Hulet}
S. E. Pollack, D. Dries, M. Junker, Y. P. Chen~F, T. A. Corcovilos, and R. G. Hulet, Phys. Rev. Lett. {\bf 102}, 090402 (2009).

\bibitem{Chevy}
N. Navon, S. Piatecki, K. J. G\"unter, B. Rem, Trong Canh Nguyen, F. Chevy, W. Krauth, 
C. Salomon, arXiv:1103.4449 (2011).

\bibitem{cluster_Tan}
Interestingly, the analytic continuation of an expression given in a note 
in reference \cite{Tan1}
(note number 16 estimating the large $k$ contribution of three-body clusters to $n(\kk$)
for a spin 1/2 Fermi gas)
to the Efimov case with a complex exponent $\gamma=\pm s_0-2$ 
correctly suggests an oscillating $1/k^5$ subleading term.

\bibitem{Tan2}
Shina Tan, 
Annals of Physics {\bf 323}, 2952 (2008).

\bibitem{Leyronas}
R. Combescot, F. Alzetto, and X. Leyronas, Phys. Rev. A {\bf 79}, 053640 (2009).

\bibitem{note2D}
The energy formula was also extended to two-dimensional gases in \cite{Leyronas},
in which case there is no problem since the Efimov effect does not occur.


\bibitem{general}
Note that the three-boson eigenwavefunctions for infinite scattering length
can actually be calculated exactly for any strength of the harmonic trap,
see \cite{Jonsell,Werner_trois}.

\bibitem{Jonsell}
S. Jonsell, H. Heiselberg, C. J. Pethick, Phys. Rev. Lett. {\bf 89}, 250401
(2002).

\bibitem{Werner_trois}
F. Werner, Y. Castin, Phys. Rev. Lett. {\bf 97}, 150401 (2006).

\bibitem{def_Rt}
Here $R_t$ is defined as e.g.\ in \cite{Werner_trois}: 
If one rescales the coordinates of all particles as $\rr_i\to
\lambda \rr_i$, with $\lambda\to 0^+$, the three-body wavefunction diverges 
as $\sin[|s_0|\ln (R/R_t)]/R^2$, with a scaling invariant prefactor and
an hyperradius $R$ defined in terms of
the Jacobi coordinates introduced below (\ref{eq:etat}) as $R=\sqrt{(r^2+\rho^2)/2}$.

\bibitem{Gradstein}
I. S. Gradshteyn and I. M. Ryzhik, 
{\sl Tables of integrals, series, and products} 
(Academic Press, 1994), 5th Edition, A.  Jeffrey, Editor.

\bibitem{Efimov93}
V. Efimov, Phys. Rev. C {\bf 47}, 1876 (1993).

\bibitem{WernerThese}
F. Werner, {\sl PhD Thesis} (Universit\'e Pierre et Marie Curie, Paris,
2008), URL http://tel.archives-ouvertes.fr/tel-00285587

\bibitem{astuce}
The following tricks are used in the numerical 
calculation of the integral appearing in (\ref{eq:qveb}). 
The integral is split in $\int_0^{k_{\rm max}} + \int_{k_{\rm max}}^{+\infty}$.
A linear scale is used to discretized $k/\kappa_0$ in between 0
and 10. For $k$ in between 10 and $k_{\rm max}$ ($k_{\rm max}$ is either 1000 or 5500), $\ln(k/\kappa_0)$
is discretized on a linear scale. The integral from $k_{\rm max}$ to infinity
is estimated with the formula $\delta n(k)\simeq -36/(k/\kappa_0)^6$, an approximate asymptotic
expression that was suggested by the numerical results 
over the range $100<k/\kappa_0<1000$.
A simple test of this overall procedure is to check the normalization of $n(\kk)$. It is found that the
numerical calculation gives the correct normalization factor within a 
$\simeq 10^{-6}$ relative error.
We note that calculating $\delta n(k)$ with a one percent error for $k=5500$
requires a calculation of $n(\kk)$ with a relative error $\simeq 2\times 10^{-10}$.

\bibitem{platter_efim}
E. Braaten, Daekyoung Kang, L. Platter, 
Phys. Rev. Lett. {\bf 106}, 153005 (2011).


\end{thebibliography}
\end{document}